# Motion of a ball rolled over a shallow step

Keith Zengel[1] and Laruen Boehnert[1]
1. Wentworth Institute of Technology, Boston, MA. email: zengelk@wit.edu

## Introduction

A ball rolled over a shallow step will experience an increase in velocity along the direction perpendicular to the step. This causes a deflection in the ball's trajectory. In this paper we derive the equations that describe the motion of a ball rolled over a shallow step and present the results of our experimental test. This simple demonstration can be used in any classroom where the physics teacher has access to a ball and a stack of papers. Prior work has shown that a ball rolled over an edge can maintain its speed, as is commonly assumed, but it can also experience an increase or even decrease in speed.[1,2] The ball can either roll without slipping while it is in contact with the edge, or else begin to slip before it leaves the edge.[3] In this paper we will consider the case where the ball rolls without slipping the entire time it is in contact with the step edge, then contacts a lower platform. We work with shallow step heights relative to the radius of the ball so that the motion of the ball is easy to observe at all times, and so that the ball does not bounce when it encounters the lower platform. These shallow step heights mean that we can assume the ball does not slip as it moves over the edge.

## Theoretical Model

A diagram of the ball on the edge is shown in Fig. 1a. Following Beeken, we can start by writing the equation for the total energy of the system while the ball is on the upper platform and while the ball is on the edge,

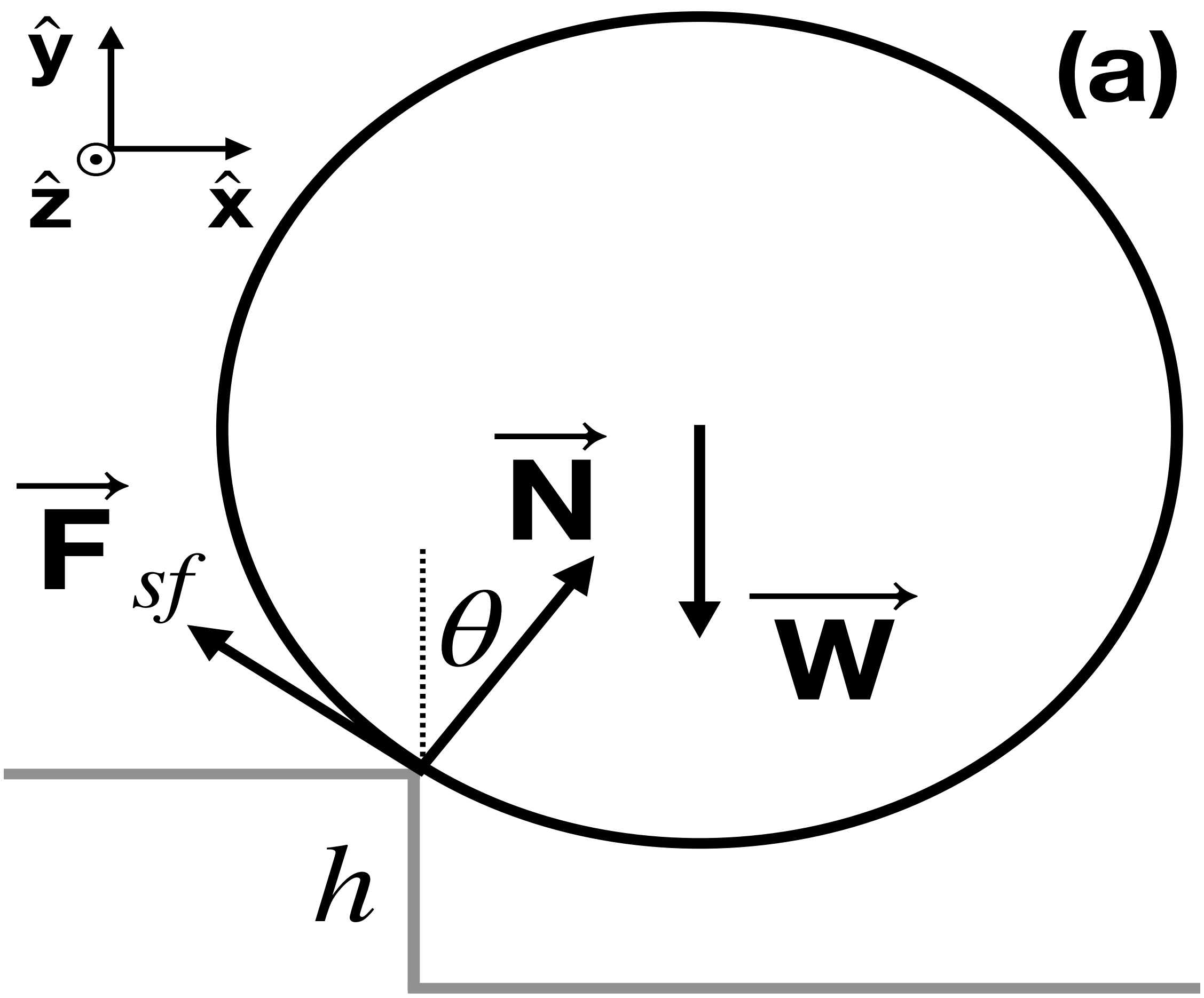

ŷ
ẑ
x̂
(a)
$\vec{\mathbf{N}}$
$\vec{\mathbf{F}}_{sf}$
$\theta$
$\vec{\mathbf{W}}$
$h$

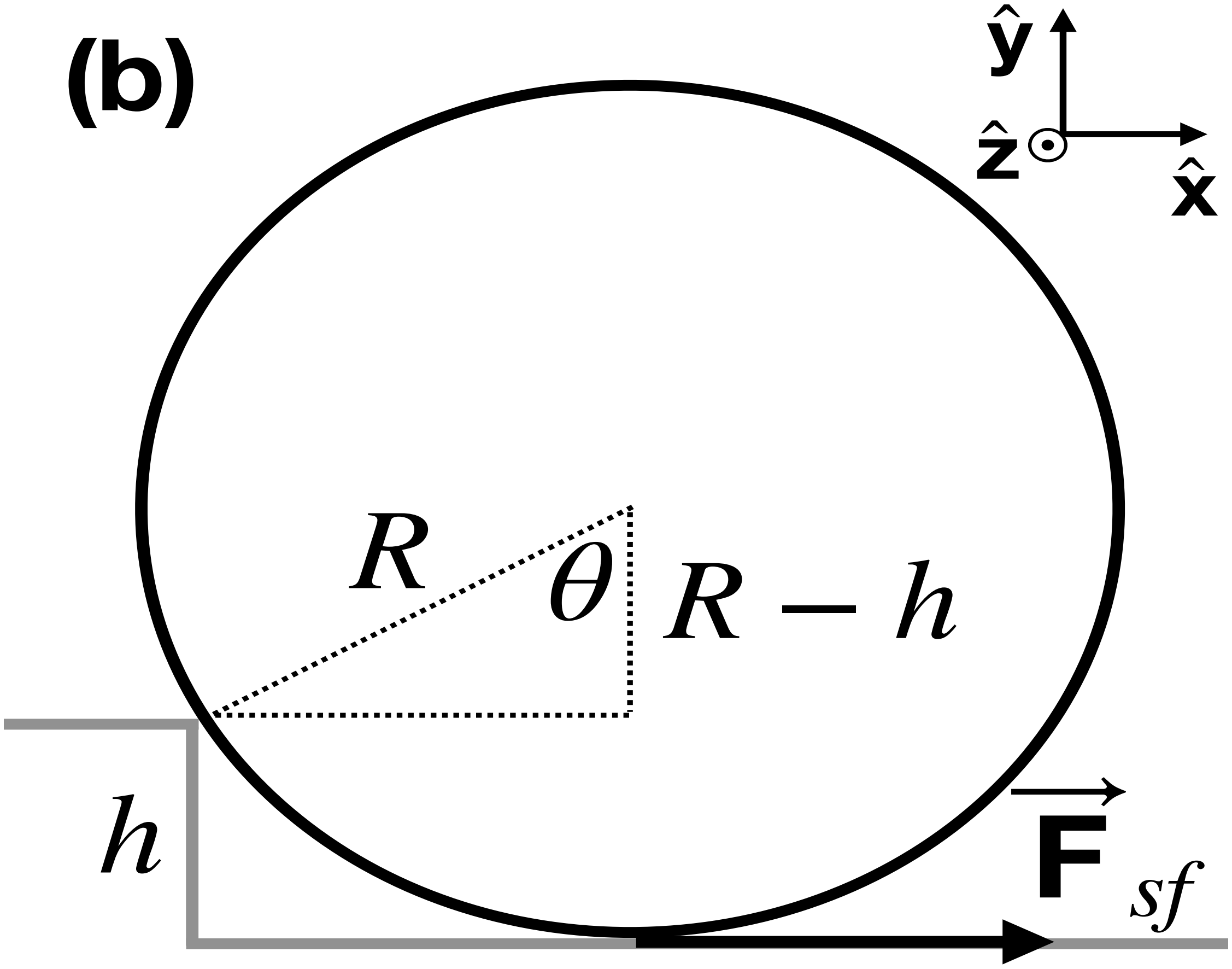

FIG. 1: A diagram of the system when the ball is in contact with the edge (a) and when the ball makes contact with the lower platform (b).

---

$$E = \frac{1}{2}I\omega_i^2 + \frac{1}{2}mv_i^2 + \frac{1}{2}mv_z^2 + mgR = \frac{1}{2}I\omega_f^2 + \frac{1}{2}mv_f^2 + \frac{1}{2}mv_z^2 + mgR\cos\theta, \tag{1}$$

where $v$ is the magnitude of the center of mass velocity of the ball in the $x$-$y$ plane, $v_z$ is the $z$-component of velocity, $\omega$ is the angular velocity of the ball along the $z$-direction, $R$ is the radius of the ball, and $I$ is the moment of inertia of the ball. Note that we have assumed the radius of the edge is much smaller than $R$. From Fig. 1a we see that no forces act in the $z$ direction, meaning $v_z$ remains constant. The condition for rolling without slipping can be written as

$$v + \omega R = 0, \tag{2}$$

and we can write the moment of inertia about the center of mass of the ball as

$$I = \beta m R^2. \quad (3)$$

Using these two identities, we can simplify Eq. 1 to find

$$v_f^2 = v_i^2 + \frac{2gR(1 - \cos\theta)}{1 + \beta}. \quad (4)$$

The equation for the net force along an axis running from the point of contact to the center of the ball is

$$-\frac{mv^2}{R} = N - mg\cos\theta. \quad (5)$$

The ball will disconnect from the edge when the normal force becomes zero and

$$\cos\theta_d = \frac{v^2}{gR}. \quad (6)$$

where $\theta_d$ is the angle at which the ball disconnects. If the ball has not disconnected from the edge before it reaches the lower platform, then it will contact the lower platform when

$$\cos\theta_e = 1 - \frac{h}{R}, \quad (7)$$

which we can see from Fig. 1b. Here $\theta_e$ is the angle of the ball when it reaches the lower platform. As long as

$$1 - \frac{h}{R} > \frac{v^2}{gR}, \quad (8)$$

the ball will reach the lower platform before disconnecting. In that case, we can substitute Eq. 7 into Eq. 4 to find

$$v_e^2 = v_i^2 + \frac{2gh}{1 + \beta}, \quad (9)$$

where $v_e$ is the velocity of the ball at the last instant it is in contact with the edge. When the ball contacts the lower platform, it experiences a normal force that exceeds the weight of the ball in magnitude. We assume that the ball does not bounce[4] and that the normal force acts to stop the ball's vertical motion instantaneously. At this time the ball also experiences a friction force acting at the bottom of the ball and pointing in the positive $x$-direction, as shown in Fig. 1b. This force occurs because the ball has an angular velocity $\omega = -v_e/R$ when it leaves the edge, but it is moving in the $x$-direction with a velocity of $v_e \cos\theta_e$. The friction force acts to increase the horizontal center of mass velocity component while also decreasing the magnitude of the angular velocity so that the ball can roll without slipping. We can model this interaction using the linear and rotational impulse equations,

$$F_{sf}\Delta t = \Delta p_x \tag{10}$$

and

$$RF_{sf}\Delta t = \Delta L, \tag{11}$$

where $p$ and $L$ are the linear and angular momenta of the ball. At the instant the ball contacts the lower platform, it has an $x$-direction linear velocity of $v_e \cos\theta$ and a $z$-direction angular velocity of $\omega = -v_e/R$. Once the ball is rolling without slipping on the lower platform, it has an $x$-direction linear velocity of $v_{xf}$ and a $z$-direction angular velocity of $\omega = -v_{xf}/R$. Combining Eqs. 7, 10, and 11 and plugging in these velocities gives

$$v_{xf} = \left[1 - \frac{h}{(1+\beta)R}\right] v_e. \tag{12}$$

Squaring both sides and combining with Eq. 9 gives

$$v_{xf}^2 = \left[1 - \frac{h}{(1+\beta)R}\right]^2 \left[v_i^2 + \frac{2gh}{1+\beta}\right]. \tag{13}$$

The case where the ball rolls from the lower platform to the higher is described by Eq. 13, after we swap $v_i$ and $v_{xf}$ and solve for $v_{xf}$:

$$v_{xf}^2 = \frac{v_i^2}{\left[1 - \frac{h}{(1+\beta)R}\right]^2} - \frac{2gh}{1+\beta}. \tag{14}$$

You can also demonstrate reflection by rolling the ball toward the edge from the lower platform with

$$v_i^2 < \left[1 - \frac{h}{(1+\beta)R}\right]^2 \frac{2gh}{1+\beta}. \tag{15}$$

In this case the ball will not have enough kinetic energy to climb to the upper platform, and will instead leave the edge with the same parallel component of velocity $v_z$ and a perpendicular velocity of $v_{xf} = -v_i$. The result in that the angle of incidence relative to the edge is the same as the angle of reflection, as expected. Alternatively, if the ball disconnects from the edge before reaching the lower platform, we find, as in Beeken's Eq. 11, that the ball leaves the edge with

$$v_{ed} = \frac{1}{gR}\left[\frac{1+\beta}{3+\beta}v_i^2 + \frac{2gR}{3+\beta}\right]^{3/2}, \tag{16}$$

where $v_{ed}$ is the horizontal component of the disconnect velocity. In this case the ball comes into contact with the ground with an $x$-direction linear velocity of $v_{ed}$ and a $z$-direction angular velocity of $\omega = -v_{ed}/R$. Once the ball is rolling without slipping on the lower platform it has an $x$-direction linear velocity of $v_{xd}$ and a $z$-direction angular velocity of $\omega = -v_{xd}/R$. Combining Eqs. 6, 10, and 11 and plugging in these velocities gives

$$v_{xd} = \frac{\beta}{(1+\beta)}\left[\frac{1+\beta}{3+\beta}v_i^2 + \frac{2gR}{3+\beta}\right]^{1/2} + \frac{1}{gR(1+\beta)}\left[\frac{1+\beta}{3+\beta}v_i^2 + \frac{2gR}{3+\beta}\right]^{3/2}. \tag{17}$$

## Experimental Results

We performed our measurements using only a small bearing ball with a diameter of 2.5 cm, two lab jacks, a level, calipers, and a camera mount. The experimental procedure was as follows. We placed two lab jacks next to each other on a level tabletop then used the level to ensure that the surface of both jacks were horizontal. We kept one jack at constant height $h_1$ and varied the other in steps of $h_1 - n * 1$ mm. The height difference $h$ was given by the difference of jack heights, which we measured with the calipers. In order to obtain uncertainties on the jack heights, we measured the height of each jack at ten different locations on different sides of the jack. For each height difference, we rolled the bearing ball over the lab jack edge several times from different angles relative to the edge. This allowed us to sample several ingoing $x$-direction velocities for each height difference (while the $z$-direction velocities remain constant throughout each trajectory). The camera was mounted above the edge. We ensured that the camera faced vertically down using a plumb line. A sketch of the experimental setup is shown in Fig. 2.

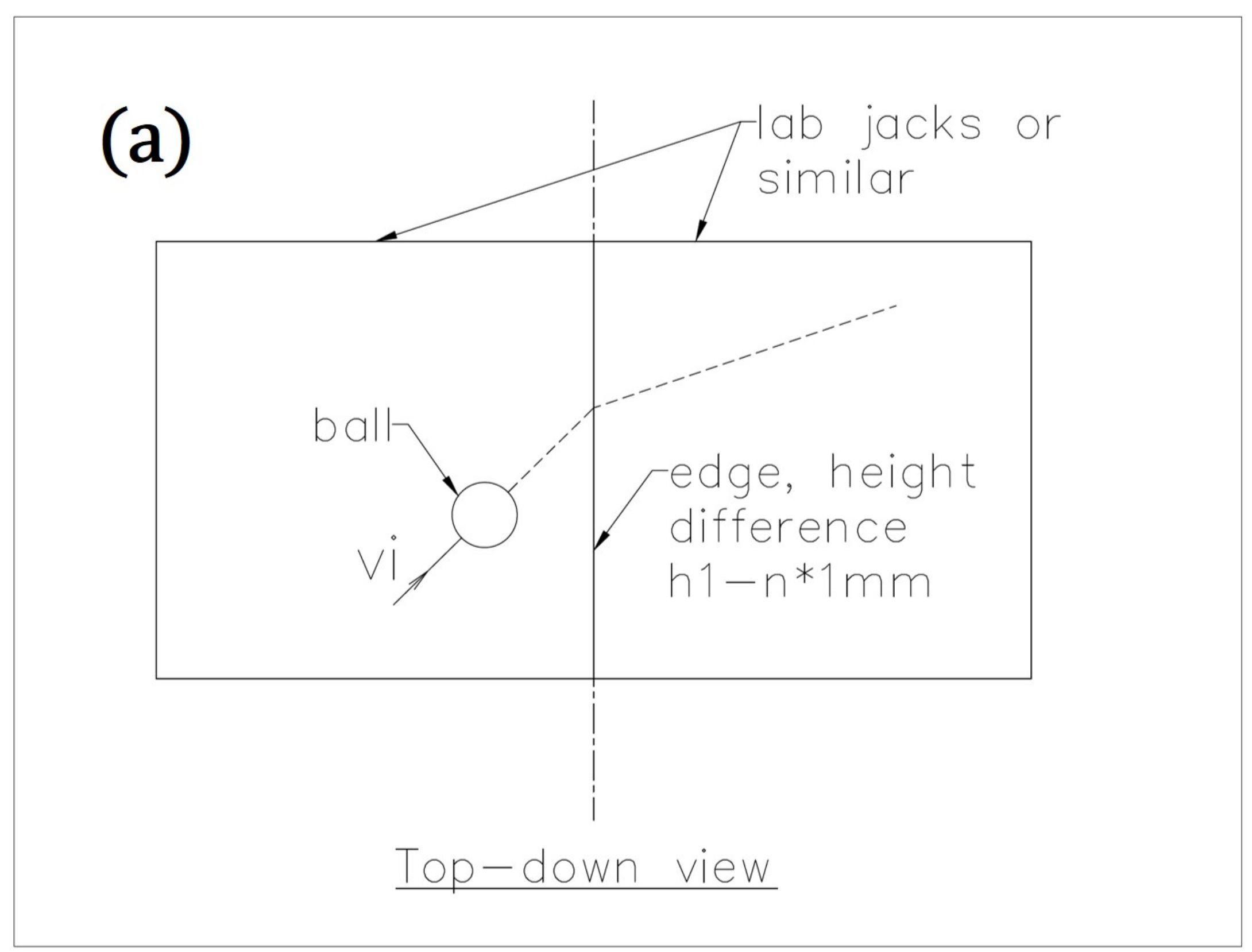
(a)
lab jacks or
similar
ball
vi
edge, height
difference
h1−n*1mm
Top−down view

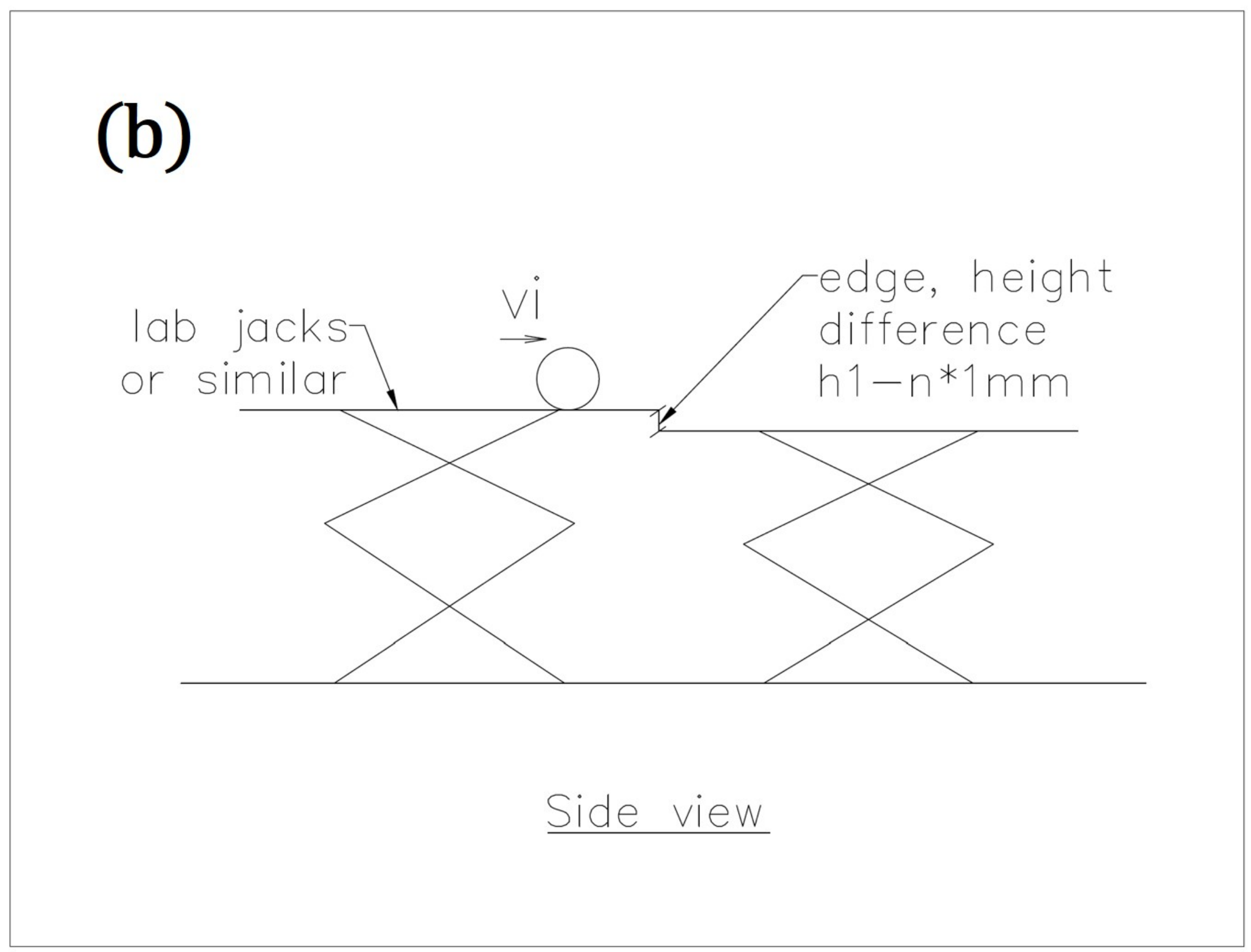

FIG. 2: A sketch of the experimental setup shown from the top-down view (a) and the side view (b).

---

Finally, we transferred the video files to Tracker[5] for analysis. A screenshot from Tracker[5] in Fig. 3 shows the path of the ball as it moves from the upper platform to the lower platform.

---

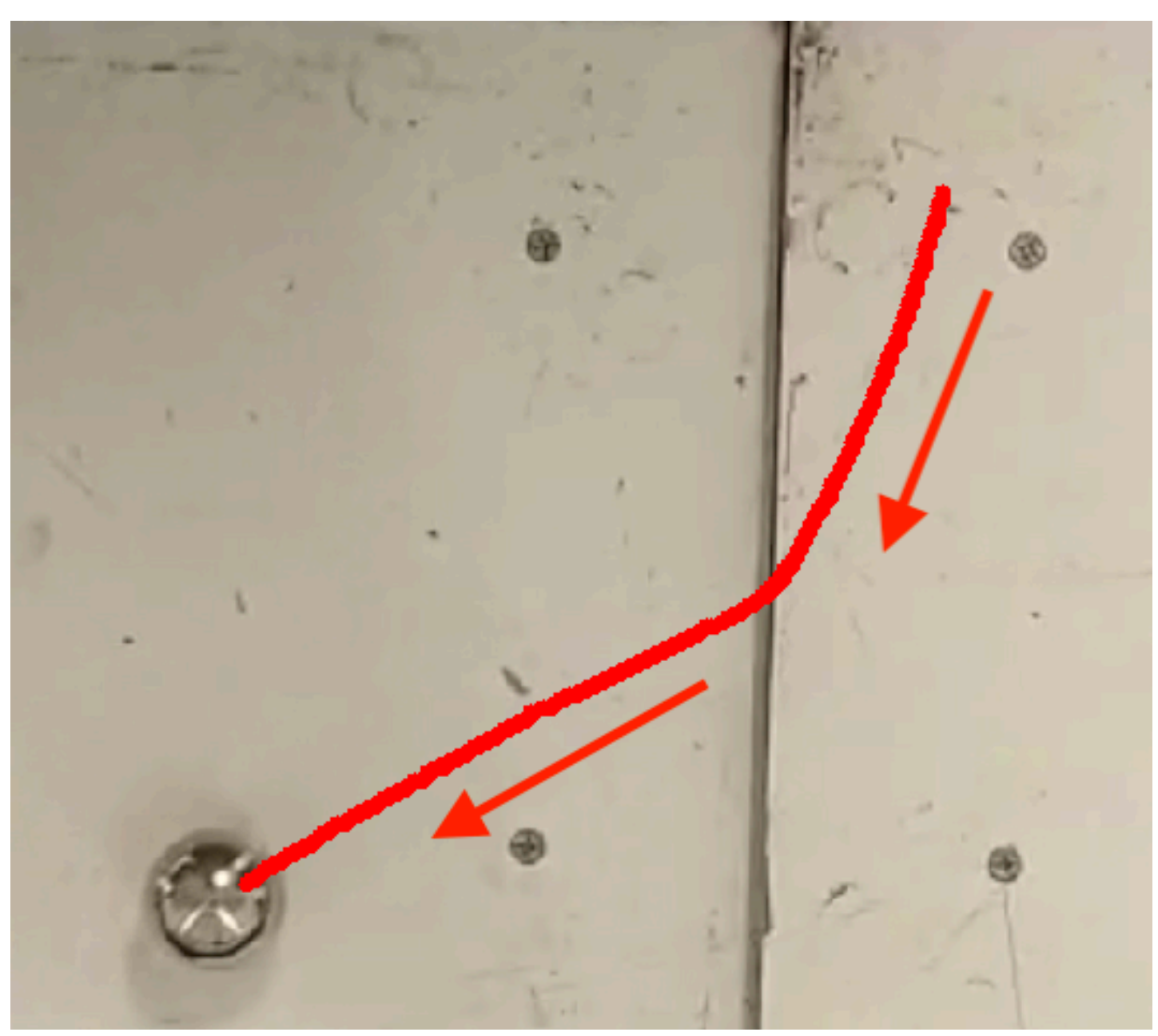

FIG. 3: A screenshot from Tracker showing the tracked path of the ball as it moves from the upper platform (right) to the lower platform (left).

---

Results for height differences of 1 mm, 3 mm, 4 mm, and 5 mm are shown in Fig. 4, along with best fit results. Fit uncertainties were obtained using a $\chi^2$ (chi-squared) fit technique. The theoretical slope and intercept for the data are given by Eq. 13 as

---

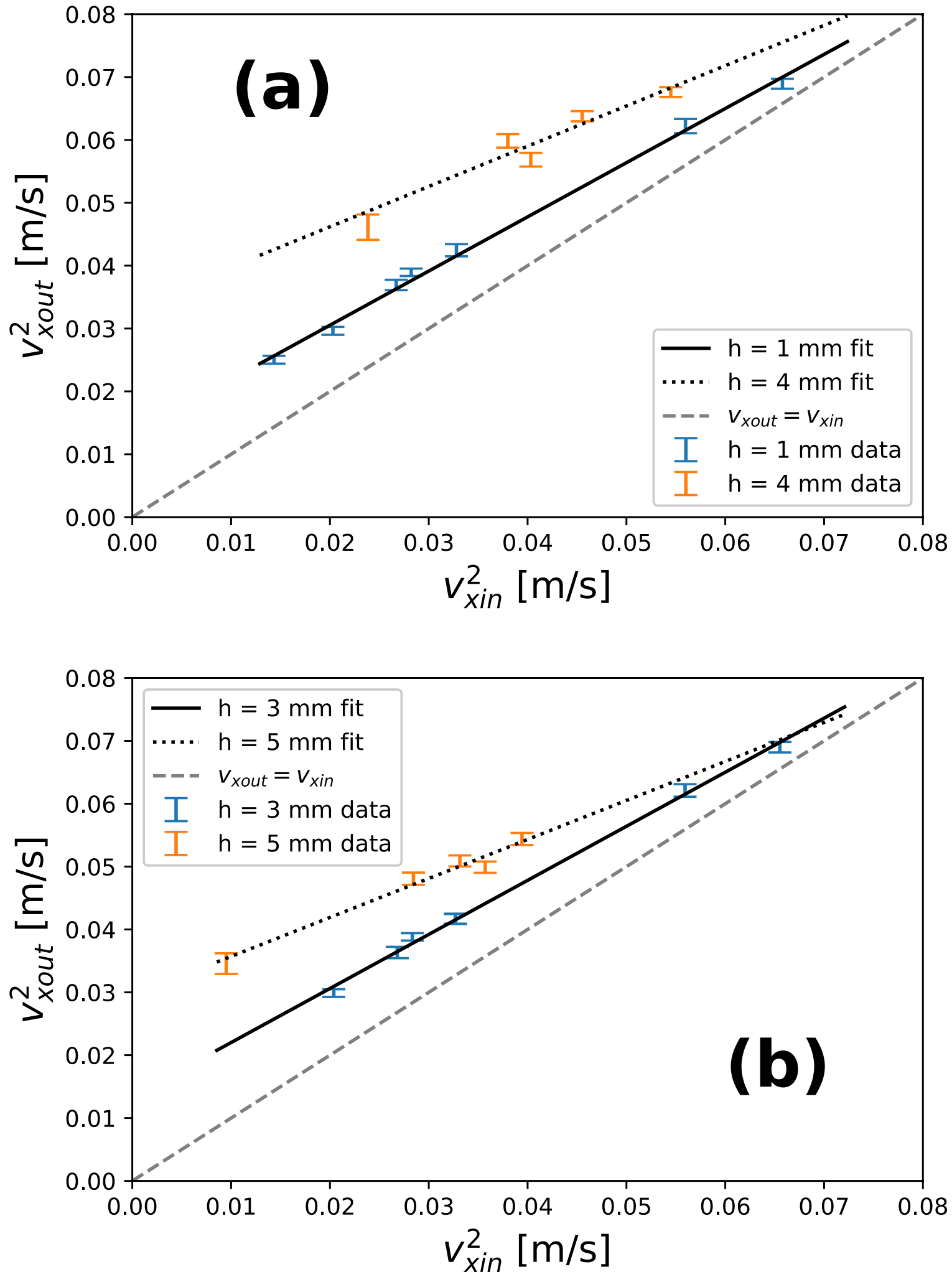

FIG. 4: The outgoing vs. ingoing squared $x$-component velocities of the ball for height differences of 1 mm and 4 mm (a) and 3 mm and 5 mm (b). Best fit lines are shown for each height difference. The $v_{\text{out}} = v_{\text{in}}$ line has been included to make the gain in velocity easier to see.

---

$$A_{\text{p}} = \left[1 - \frac{h}{(1+\beta)R}\right]^2 \tag{18}$$

and

$$B_{\text{p}} = \left[1 - \frac{h}{(1+\beta)R}\right]^2 \left[\frac{2gh}{1+\beta}\right], \tag{19}$$

where

$$v_{xf}^2 = A_p v_i^2 + B_p. \quad (20)$$

A summary of the predicted and measured values are given in Table I. The dominant uncertainty is on the height of the lab jacks, which varies around the edges. Agreement between the theoretical model and measurement is generally within one standard deviation. Previous results[6] suggest that a steel ball rolling on a steel incline will begin to slip on inclines at angles of roughly 25° above the horizontal. In our experiments the maximum inclination angle of a ball occurs with $h = 5$ mm, which according to Eq. 7 means an angle of roughly 11°. Therefore it seems reasonable to conclude the balls in our trials roll over the edge without slipping. When the ball is rolled with greater speeds, or when the height differences between lab jacks are greater, the ball will disconnect before contacting the lower platform. In that case the final speed of the ball is given by Eq. 17, which has four fit-parameters. We found generally good agreement between the model and our measurements, but this was primarily due to the large uncertainties on the four fit parameters, which meant our measurements were consistent with nearly every plausible model.

TABLE I: The predicted and measured slope and intercept values for various height differences. Uncertainties on the predicted values come from the uncertainties on the lab jack heights.

| Height Diff. [ mm ] | Predicted slope $A_m$ | Measured slope $A_p$ | Predicted intercept $B_p$ [ $\mathrm{m}^2/\mathrm{s}^2$] | Measured Intercept $B_m$ [ $\mathrm{m}^2/\mathrm{s}^2$] |
|---|---|---|---|---|
| $1.0 \pm 0.7$ | $0.91 \pm 0.17$ | $0.86 \pm 0.04$ | $0.010 \pm 0.008$ | $0.013 \pm 0.0014$ |
| $3.0 \pm 0.7$ | $0.70 \pm 0.14$ | $0.86 \pm 0.05$ | $0.028 \pm 0.008$ | $0.0134 \pm 0.0018$ |
| $4.0 \pm 0.7$ | $0.61 \pm 0.13$ | $0.64 \pm 0.14$ | $0.033 \pm 0.007$ | $0.033 \pm 0.006$ |
| $5.0 \pm 0.7$ | $0.52 \pm 0.12$ | $0.62 \pm 0.15$ | $0.036 \pm 0.007$ | $0.030 \pm 0.005$ |

## Conclusion

This simple demonstration can performed in any classroom where the teacher is able to roll a ball from one lab jack to another, or from a small book to the surface of a table. The experiment is also suitable for introductory physics labs, either as its own lab or as a modification of the lab suggested by Beeken.